\documentclass[aps,prb,twocolumn,showpacs,floatfix]{revtex4-1}

\usepackage{graphicx,color}
\usepackage{amsfonts}
\usepackage[figuresright]{rotating}  
\usepackage{amssymb}
\usepackage{amsmath}
\usepackage{mathtools}
\usepackage{psfrag}
\usepackage{subfigure}
\usepackage{multirow}
\usepackage{tabularx}
\usepackage{textcomp}
\usepackage{units}
\usepackage{hyperref}
\hypersetup{
 pdfnewwindow=true, colorlinks=true,
 linkcolor=blue, anchorcolor=blue,
 citecolor=blue, filecolor=blue,
 menucolor=blue, urlcolor=blue}

\def\beq{\begin{eqnarray}}
\def\eeq{\end{eqnarray}}

\renewcommand{\v}[1]{\ensuremath{\mathbf{#1}}} 
 
\let\baraccent=\= 
\renewcommand{\=}[1]{\stackrel{#1}{=}} 
	
\makeatletter

\begin{document}
\title{Strong second harmonic generation in two-dimensional ferroelectric IV-monochalcogenides}

\author{Suman\ \surname{Raj Panday}}
\affiliation{Department of Physics, Kent State University, Kent,
Ohio, 44242, USA}

\author{Benjamin\ \surname{M. Fregoso}}
\affiliation{Department of Physics, Kent State University, Kent,
Ohio, 44242, USA}

\begin{abstract}
The two-dimensional ferroelectrics GeS, GeSe, SnS and SnSe are expected to have large spontaneous in-plane 
electric polarization and enhanced shift-current response. Using density functional methods, we show 
that these materials also exhibit the largest effective second harmonic generation reported so far. It can reach 
magnitudes up to $10$ nm/V which is about an order of magnitude larger than that of prototypical GaAs. To rationalize this 
result we model the optical response with a simple one-dimensional two-band model along the spontaneous polarization direction. 
Within this model the second-harmonic generation tensor is proportional to the shift-current response tensor. 
The large shift current and second harmonic responses of GeS, GeSe, SnS and SnSe make them promising 
non-linear materials for optoelectronic applications.
\end{abstract}

\maketitle

\section{Introduction}
\label{sec:intro}
The second harmonic generation (SHG) is one of the most important non-linear optical responses 
in semiconductor physics\cite{Boyd2008}. Common applications include frequency doublers and surface 
probes where the extreme sensitivity to local crystal symmetry is exploited. The SHG research has a 
long history dating back to the 60's. It has been investigated extensively in 
bulk~\cite{Aspnes1972,Levine1990,Sipe1993,Aversa1995,Sipe2000,Luppi2016} and more recently two-dimensional (2D) 
materials~\cite{Kumar2013,Li2013,Kim2013,Janisch2014,Malard2013,Zhou2015,Attaccalite2015}.
The SHG is non-vanishing only in materials that lack inversion symmetry. These can be polar with a 
finite electric polarization or non-polar. To our knowledge the SHG has not been investigated in 
2D ferroelectrics. In 2016 ferroelectricity was realized in single-layer SnTe~\cite{Chang2016},
motivating our study of optical response in lower dimensional ferroelectrics. The advent of 2D ferroelectrics 
provides a new playground for experimentalists and theorists in the search for new optical phenomena where 
dimensionality and ferroelectricity play an important role. 

Recent studies of single-layer GeS, GeSe, SnS and SnSe (hereafter referred as MX) show that they
are 2D ferroelectrics. These materials are predicted to have very large in-plane spontaneous electric 
polarization~\cite{Mehboudi2016,Wu2016,Wang2017,Rangel2017} and large shift-current 
response~\cite{Rangel2017,Cook2017,Fregoso2017}. Variation of these structures such the 
$\beta-$GeSe also show promising transport properties too~\cite{Rohr2017}. 
The shift current~\cite{Baltz1981,Sturman1992,Sipe2000,Young2012,Tan2016a} is the first non-vanishing 
contribution to dc current in ferroelectrics. Similar to the SHG, the shift current is quadratic 
in the electric field and hence only present in materials that lack inversion symmetry. Intuitively, the 
electron wavepacket jumps from one atom to another when absorbing a photon. It requires quantum coherence 
but does not require the medium to be inhomogeneous. Importantly, the SHG susceptibility tensor diverges at shift 
current states\cite{Sipe2000}, and hence is it interesting to investigate whether the SHG is also large in 
MX or if it is related to its large shift current.

In this work we answer these questions affirmatively. We calculate the SHG in MX using \textit{ab}-\textit{initio} 
density functional theory (DFT). We find that the response is larger than prototypical non-linear semiconductor 
GaAs by an order of magnitude. We model the optical response along the polar axis with a 
two-band approximation which allows us to disentangle the contributions to the SHG susceptibility. Within 
this approximation the imaginary part of the SHG response is proportional to the shift current tensor
and the real part is proportional to the shift vector~\cite{Baltz1981,Sturman1992,Sipe2000}. 
Since the shift current is large in MX~\cite{Rangel2017}, we expect large SHG, consistent with our DFT 
calculations. The model further predicts that the integral of the imaginary part of the 
SHG tensor (along the polar axis) times the frequency vanishes. This prediction is tested 
against multiband DFT calculations for MX finding good agreement. 

In section~\ref{sec:methods} we give the details of the numerical computations and
in Sec.~\ref{sec:num_results} we present the DFT results for the SHG susceptibility. In Sec.~\ref{model_shg}
we construct a two-band approximation and compared it with our DFT results. 
We conclude in Sec.~\ref{sec:conclusions}.
 
\begin{figure}[]
\subfigure{\includegraphics[width=.5\textwidth]{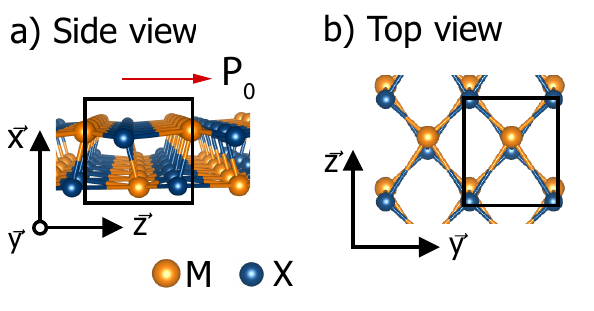}}
\caption{Crystal structure of single-layer MX=GeS, GeSe, SnS, SnSe. The black square indicates the unit cell. 
The spontaneous electric polarization is in-plane $\v{P}_0 =P_0 \v{z}$}
\label{fig:structure GeS}
\end{figure}

\begin{figure*}[]
\subfigure{\includegraphics[width=.95\textwidth]{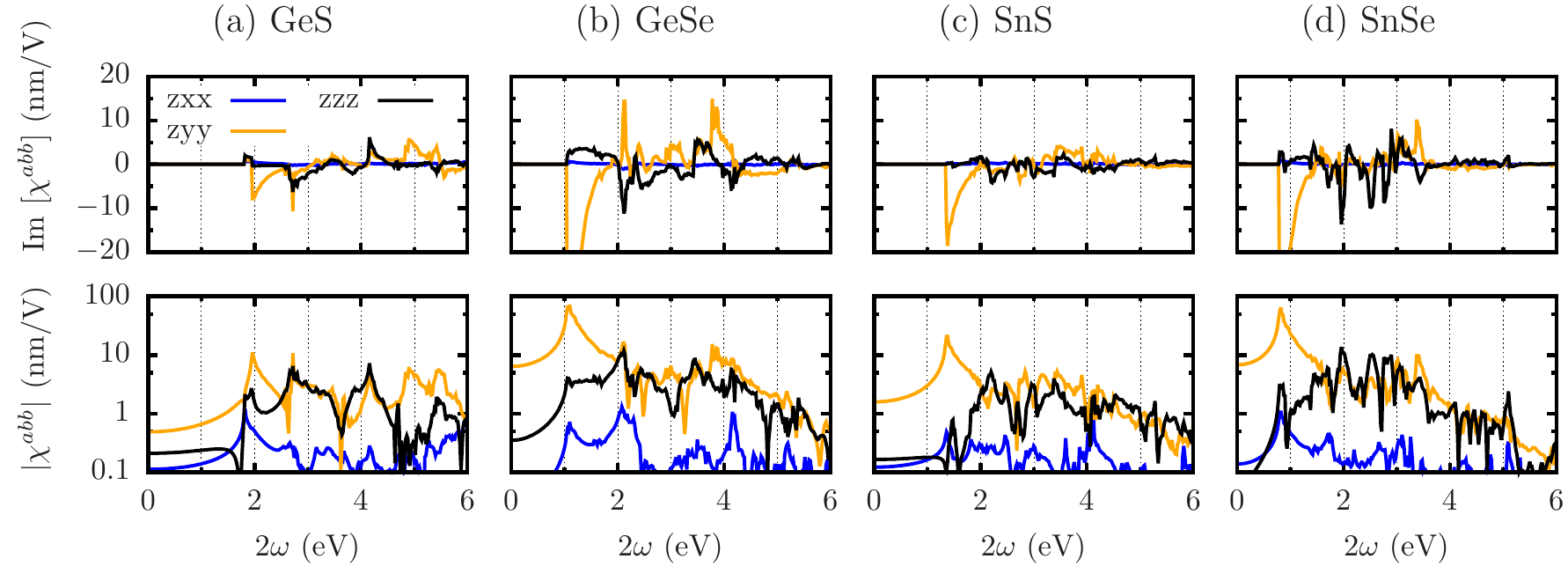}}
\caption{(a)-(d) Second harmonic generation (SHG) tensor $\chi_{2}^{abb}(-2\omega,\omega,\omega)$ 
for single-layer monochalcogenide MX=GeS, GeSe, SnS, SnSe as a function of outgoing photon 
frequency $2\omega$. The response can reach $\sim 10$ nm/V in the visible range of frequency 
which makes them promising for opto-electronic applications.}
\label{fig:shg}
\end{figure*}

\section{Methods}
\label{sec:methods}
We use density functional theory (DFT) as implemented in the ABINIT~\cite{Gonze2009} computer package with the generalized 
gradient approximation to the exchange correlation energy functional within the 
Perdew-Burke-Ernzerhof functional~\cite{Perdew1996}. We use Hartwigsen-Goedecker-Hutter norm conserving 
pseudopotentials~\cite{Hartwigsen1998} available in the ABINIT website~\cite{abinit_website}. We use an energy cut-off of 40 
Hartrees to expand the plane waves basis set. To model the slabs we use supercells
with 15 \AA~ along the non-periodic direction, which includes about 10 \AA~ of vacuum. 
To calculate the SHG we include 20 valence and 30 conduction bands, together they accounts for all 
allowed transitions up to 6 eV; we use a mesh of $70\times 70$ $\v{k}$-points along the
periodic slab directions. 

In periodic-cell calculations, the response is integrated over the three-dimensional (3D) Brillouin zone (BZ). 
In particular, supercells calculations include a vacuum region used to simulate 2D slabs. This contribution 
must be subtracted from the total response. To extract the effective response of a single layer we factor 
the response per unit length perpendicular to the slab and multiply by an effective thickness of layer. 
The procedure amounts to scale the supercell numerical results by a factor $L/d$, where $L$ is 
the supercell lattice parameter perpendicular to the slab and $d$ is the effective thickness of the layer. Here we assume 
reasonable effective slab widths of $d=$2.56, 2.61, 2.84 and 2.73 \AA,~ for GeS, GeSe, SnS and SnSe, respectively. 
Once the ground-state wave function and energies are computed, we compute the SHG susceptibility $\chi_2$, as implemented 
TINIBA\cite{tiniba} which is based on the analysis of reference~\onlinecite{Sipe2000}. The sum over $\v{k}$-points 
is made using the interpolation tetrahedron method\cite{bloch-tetra}. Our calculated band structures agree 
with previous reports and our method of calculating SHG reproduces that of GaAs reported in the 
literature~\cite{Cabellos2009}.


\section{Results}
\label{sec:num_results}
The electric polarization in materials can be descried as a power series in 
the electric field~\cite{Boyd2008},
\begin{align}
\v{P}=\v{P_0}+\chi_1 \v{E} + \chi_2 \v{E}^2 + \chi_3 \v{E}^3 + \cdots,
\end{align}
Where $\v{P}_0$ is the electric polarization in the absence of fields, $\chi_1$ is the linear 
susceptibility and $\chi_2$, $\chi_3$, etc. are the non-linear susceptibilities; $\v{E}$ is the 
locally-averaged macroscopic electric field (local-field effects are not included in this work). 
For MX, $\v{P}_0=P_0 \v{z}$ is parallel to the slab as shown in Fig.~\ref{fig:structure GeS} and 
can be as large as 1.9 C/m$^2$~\onlinecite{Rangel2017}. 
For a monochromatic electric field, $E^a = E^{a}(\omega)e^{-i\omega t}$ + c.c., the second order 
polarization can be expressed as, 
\begin{align}
 P^{a}_2(t)=
\chi^{abc}_2 (-2\omega;\omega,\omega) E^{b}(\omega) E^{c}(\omega) e^{-i 2\omega t} + \textrm{c.c.},
\label{eqn:chi_2}
\end{align}
where $a,b$ and $c$ are Cartesian components along $x,y,z$ direction and summation 
over repeated indices is implied. The space group of bulk MX is $Pnma$, which contains 
a center of inversion and hence has zero bulk spontaneous polarization. The atoms in the 
conventional cell are arranged  in two weakly coupled layers, each with opposite in-plane 
polarizations. When one of the layers is removed, the 
resulting structure lacks inversion symmetry and has large in-plane spontaneous electric 
polarization~\cite{Wu2016,Wang2017,Rangel2017}.The single-layer of MX has point group 
$mm2$ and so the only non-zero components of $\chi_2$ 
are $zxx$, $zyy$, $zzz$, $yyz$, $xzx$, as well as the components obtained from exchanging 
the last two indices. 

\begin{figure*}[]
\subfigure{\includegraphics[width=0.95\textwidth]{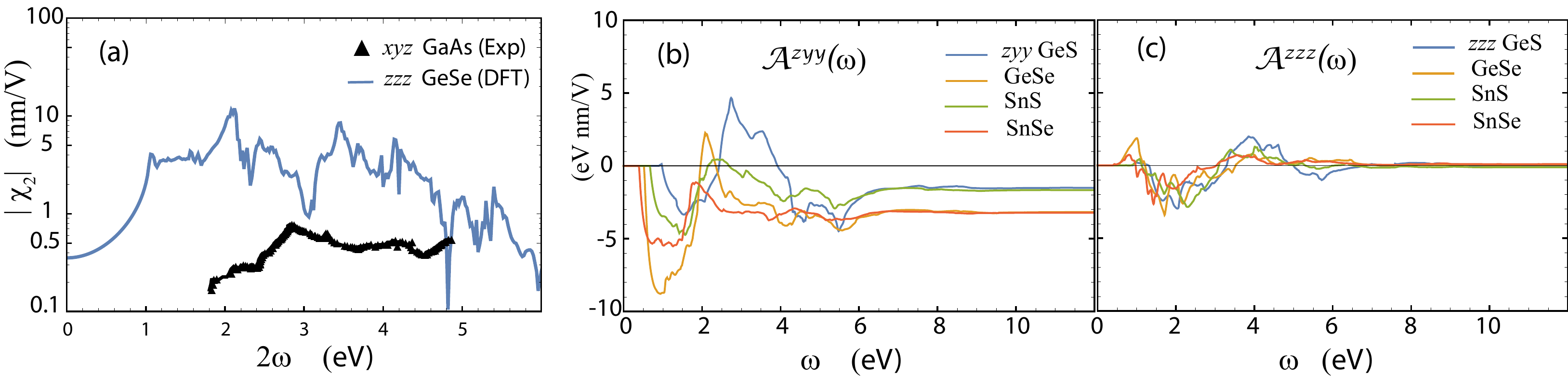}}
\caption{(a) Comparison between experimental SHG tensor in GaAs(001) from~\onlinecite{Bergfeld2003} 
and our density functional theory (DFT) calculation for GeSe. The SHG in GeSe is an order of magnitude
larger than in GaAs.(b)-(c) Integral of imaginary part of SHG defined in Eq.~\ref{eq:shg_sumrule_dft} 
for the $zyy$ (b) and $zzz$ (c) components. As $\omega$ increases $\mathcal{A}^{zzz}$ vanishes 
within numerical precision but  $\mathcal{A}^{zyy}$ does not. 
Hence the sum rule Eq.~\ref{eq:shg_int} is non-trivially satisfied.}
\label{fig:shg_integral}
\end{figure*}

In Fig.~\ref{fig:shg} we show our DFT results for the imaginary and absolute part of $\chi_2$ 
for MX=GeS, GeSe, SnS, and SnSe. Only the components giving rise to a current along the polar 
axis with linear polarization are shown. The other components are much smaller with the exception 
of $yyz$ which is of the same order as the $zzz$ component. Note that the effective $|\chi_2|$ can reach values up to $10$ nm/V over 
a large frequency range including the visible frequency regime ($1.5$-$3$ eV). 
In fact, it is larger than that of prototypical semiconductor GaAs~\cite{Bergfeld2003}, which 
can reach up to $0.8$ nm/V, see Fig.~\ref{fig:shg_integral}(a). Another 
interesting feature is that the strong in-plane anisotropy of $\chi_2$ in MX, e.g., 
$|\chi_{2}^{zyy}|$ is generally larger than $|\chi_{2}^{zzz}|$. In table ~\ref{table:shg_2d}, 
we compare the reported values of the SHG of other 2D materials studied so far. Even though 
these materials break inversion symmetry they have a rotational 3-fold symmetry which 
prevents them from having a polar axis. As a consequence they have zero electric polarization, 
except for MX studied in this work. Indeed, MX has the largest SHG reported so far. 
   
\begin{table}[]
\begin{center}
 \begin{tabular}{||c c c c c||} 
 \hline
 Monolayer & ~$|\chi_2|$(nm/V)~ & $ \hbar\omega$ (eV) & Ref. & $P_0$ (C/m$^2$) \\ [0.2ex] 
 \hline\hline
 MX & 10 & 0.8-4  & present &  0.6-1.9~\cite{Rangel2017}  \\
 \hline
 WS$_2$ & 4.5 & 3 & ~\onlinecite{Janisch2014}(exp.) &0 \\
 \hline
 GaSe & 2.4  & 1  & ~\onlinecite{Zhou2015} (exp.) & 0 \\ 
 \hline
 SiC & 0.6  & 1.5   & ~\onlinecite{Attaccalite2015}(th.)&0 \\
 \hline
 MoS$_2$ & $0.16$ & 1.5  & ~\onlinecite{Li2013}(exp.) & 0 \\ 
 \hline
 ZnO,GaN & 0.08  & 1.5  & ~\onlinecite{Attaccalite2015}(th.)&0 \\ 
\hline
 h-BN & 0.001 & 1.5   & ~\onlinecite{Li2013}(exp.)&0 \\ [1ex]
\hline	
\end{tabular}
\end{center}
\caption{Reported second harmonic generation (SHG) tensor for various 2D materials. 
MX stands for GeS, GeSe, SnS or SnSe. Values are meant to be order of magnitude  estimates. $\hbar\omega$ 
is incoming photon energy. A large value of SHG in MoS$_2$ is reported 
in~\onlinecite{Kumar2013} but has been recently challenged~\cite{Merano2016}.
Note that of the materials studied to date only MX has finite spontaneous electric 
polarization $P_0$~\cite{Wu2016,Wang2017,Rangel2017}. Experimental (exp.) and theoretical (th.) 
values are indicated.}
\label{table:shg_2d}
\end{table}

\section{Two-band approximation of the SHG susceptibility }
\label{model_shg}
Since the analytic expression for $\chi_2$ is not simple (even at the single-particle level) it is hard 
to disentangle the important factors contributing to the magnitude of 
$\chi_2$ or any correlations to other optical responses. However, simple two-band models such as 
the Rice-Mele model~\cite{Rice1982}, have been successful in explaining the relation between shift current and 
electric polarization~\cite{Rangel2017}. In this work, we expand this approach and model the SHG along the polar axis, 
with a two-band approximation. From the general expression for $\chi_2$ (see~\onlinecite{Sipe2000}),
setting $a,b,c=z$ and for simplicity omitting the Cartesian components we obtain,
\begin{align}
\chi_2 =  \chi'_{2} + i \chi{''}_{2},
\end{align}
with,
\begin{align}
\chi{'}_{2}  = \frac{e^3}{\hbar^2} &\int \frac{dk}{2\pi}  \frac{1}{\omega_{21}}  |r_{21}| R_{21} \nonumber \big[ 2 H_{+}(\omega_{21},2\omega) \\
& 	~~~~~~~~~~+ \frac{(\omega_{21}-2\omega)}{2\omega} H_{+}(\omega_{21},\omega) \big] 
\label{eqn:shg_real} \\
\chi{''}_{2}  = \frac{1}{ 2\omega} &\big[2 \sigma_{2}(0;-2\omega,2\omega)-\sigma_{2}(0;-\omega,\omega) \big],
\label{eqn:shg_im}
\end{align}
where $\sigma_{2}(0;-\omega,\omega)$ is the shift current tensor. 
\begin{align}
\sigma_{2}(0;-\omega,\omega) &= \frac{\pi e^3}{\hbar^2} \int \frac{dk}{2\pi} |r_{21}|^2 R_{21}\delta(\omega_{21}-\omega).
\label{eq:shift_c_tensor}
\end{align}
In these expressions both the positive and negative components of the frequency were taken into account 
and $\omega$ is positive. The shift vector,
\begin{align}
R_{21} &=	 \frac{\partial \phi_{21}}{\partial k}  + A_{22} - A_{11},
\label{eqn:shift_vector}
\end{align}
depends on the Berry connections, $A_{nm}=i\langle u_n|\partial_k|u_m\rangle$, where $n,m$ are
band indices. 
$r_{12} = v_{12}/i\omega_{12}$ is the dipole matrix element, and $v_{12}=\langle u_1|v| u_2 \rangle$ are
 the velocity matrix elements between the periodic part of the Bloch states 
$u_{m}$. $\phi_{nm}$ is the phase of the interband Berry connection,  
$A_{nm}= |A_{nm}|e^{-i \phi_{nm}}$. $H_{\pm}(\omega_{nm})$ is 
\begin{align}
H_{\pm}(\omega_{nm},\omega) &=  \frac{P}{\omega_{nm} -\omega} \pm \frac{P}{\omega_{nm} + \omega},
\end{align}
where $\hbar\omega_{21} = \hbar\omega_{2}-\hbar\omega_{1} = 2E_k$,  $E_k$ is the band dispersion and
$P$ takes the principal part of the argument. These expressions can also be obtained from Floquet 
theory~\cite{Morimoto2016}. Since $\chi{''}_i$  ($i$=1,2,..) is related to the electromagnetic energy stored 
in a dielectric~\cite{Boyd2008}, $\chi{''}_2$ is zero when there is no energy absorption. 
From Eq.~\ref{eqn:shg_im} we see that $\chi{''}_2=0$ for $2\hbar\omega < E_g$, because  
two photons of energy at least $\hbar\omega=E_g/2$ can be absorbed. The real part however can be non-zero below the gap
energy due to virtual transitions.

The imaginary part $\chi{''}_2$ is proportional to the difference of two shift current tensors at the first and 
second harmonic of the incoming photon frequency. From previous studies we know the shift 
current tensor~\cite{Cook2017,Tan2016a} has sharp 
onset at the band edges. Hence $\chi{''}_2$ has two sharp peaks at $\hbar\omega=E_g$ and $2\hbar\omega=E_g$.
Barring a fortuitous cancellation between the peaks, a large shift current would imply 
large $\chi{''}_2$. More important, the shift current tensor in Eq.~\ref{eq:shift_c_tensor} depends
on a gauge-invariant matrix elements $|r_{12}|^2 R_{12}$ and the 
density of states (DOS). Usually, these contributions cannot be disentangled and the shift current has a complex
dependence of each of them~\cite{Young2012,Tan2016a}. In 2D, the situation is different. The DOS 
is approximately constant and hence the shift current (and $\chi_2$) are determined by the shift 
vector and velocity matrix elements~\cite{Fregoso2017}. This means that, for materials  
with similar DOS, the one with larger spontaneous electric polarization have larger shift current and hence stronger SHG. 
As shown above our DFT calculations of the SHG in MX are consistent with this picture. 

\subsection{Sum rule} 
Integrating the imaginary part of $\chi_2$, as given in
Eq.~\ref{eqn:shg_im} we obtain,
\begin{align}
\int_{0}^{\infty} d\omega~ 2\omega~ \chi{''}_{2}(-2\omega,\omega,\omega)  = 0,
\label{eq:shg_int}
\end{align}
where we used $r_{21} = v_{21}/i \omega_{21}$. This result is not a simple consequence of 
the oddness of $\chi^{''}_2$ under $\omega\to -\omega$ and hence it is interesting to assess 
its validity for a full-band structure calculation. To this end we define,
\begin{align}
\mathcal{A}^{zzz}(\omega)=\int_{0}^{\omega} d\Omega~ 2\Omega~ \chi^{''zzz}_{2}, 
\label{eq:shg_sumrule_dft}
\end{align}
and computed $\mathcal{A}^{zzz}(\omega)$ within DFT. We find the sum rule is mostly satisfied for 
$\omega > 7$ eV for the materials considered, as shown in Fig.~\ref{fig:shg_integral}(c). 
The sum rule is not satisfied for other components; for instance in Fig.~\ref{fig:shg_integral}(b)
we show $\mathcal{A}^{zyy}$. Hence the two-band approximation captures the 
integrated SHG response along the polar axis in MX.

\section{conclusions}
\label{sec:conclusions}
We have calculated the second harmonic generation (SHG) susceptibility of single-layer GeS, GeSe, SnS and SnSe using
density functional theory. We found that the effective 3D SHG response of these materials is larger than that of GaAs 
by an order of magnitude and is the largest reported so far. In addition, we constructed a two-band approximation 
to SHG multiband susceptibility to describe the SHG. Within this approximation we found that the imaginary part 
of $\chi_2^{zzz}$ is proportional to the difference of two shift current tensors at the first and second harmonic frequencies.
Since the shift current is large in MX\cite{Rangel2017,Cook2017}, we expect the SHG will be large too, in 
agreement with our DFT calculation. 

We left for future research the inclusion of quasiparticle effects, local fields 
and excitonic contributions. Quasiparticle and local fields effects
are expected to be small and could be well approximated within the independent-particle 
formalism~\cite{Luppi2016}. Excitonic bound states, on the other hand, are expected 
to \textit{increase} $|\chi_2|$ due to resonances at bound states. In conclusion, the strong SHG 
together with the large shift current in GeS, GeSe, SnS and SnSe make these materials
of great interest for optoelectronic applications.
 	
\section{acknowledgments}
We thank T. Rangel and C. Salazar for their help with TINIBA and ABINIT and M. Merano for kindly 
bringing to our attention reference~\onlinecite{Merano2016}. We acknowledge support from NERSC contract 
No. DE-AC02-05CH11231.

\end{document}